\documentclass[runningheads]{svmult}
\usepackage{makeidx}
\usepackage{graphicx}
\usepackage{subeqnar}
\usepackage{multicol}
\usepackage{physmubb}
\usepackage{amssymb}
\usepackage{amsmath}
\makeindex
\begin{document}

\title*{Development of a density inversion\\ in driven Granular Gases}
\toctitle{Development of a density inversion in driven Granular Gases}
\titlerunning{Development of a density inversion in driven Granular Gases}
\author{Yaron Bromberg\inst{1}
\and Eli Livne\inst{2}
\and Baruch Meerson\inst{2}}
\authorrunning{Y. Bromberg, E. Livne, and B. Meerson}
\institute{School of Physics and Astronomy, Tel Aviv University, Tel Aviv 69978, Israel \and
Racah Institute  of Physics, Hebrew University of  Jerusalem, Jerusalem 91904, Israel}
\maketitle

\begin{abstract}
Granular materials fluidized by a rapidly vibrating bottom plate
often develop a fascinating density inversion: a heavier layer of
granulate supported by a lower-density region. We employ the
Navier-Stokes granular hydrodynamics to follow a density inversion
as it develops in time. Assuming a dilute low-Mach-number flow, we
derive a reduced time-dependent model of the late stage of the
dynamics. The model looks especially simple in the Lagrangian
coordinates. The time-dependent solution describes the growth of a
density peak at an intermediate height. A transient temperature
minimum is predicted to develop in the region of the density peak.
The temperature minimum disappears at later times, as the system
approaches the steady state. At late times, the predictions of the
low-Mach-number model are in good agreement with a numerical
solution of the full hydrodynamic equations. At an early stage of
the dynamics, pressure oscillations are predicted.
\end{abstract}

\section{Introduction}

An ensemble of hard spheres in motion, that collide inelastically
and are characterized by a constant coefficient of normal
restitution, represents the simplest model of granular gas
\cite{Kadanoff}. It also provides an excellent example of
complexity. Indeed, the properties of the granular gas as a whole,
such as clustering
\cite{clustering1,clustering2,clustering3,clustering4}, come out
as non-trivial consequences of the simple algebraic laws that
govern the velocities of the binary collisions of individual
particles. Furthermore, being intrinsically far from thermal
equilibrium, granular gas defies statistical mechanics, the
cornerstone of the equilibrium science. As the result, formulating
a universally applicable continuum theory of granular gas, even in
the dilute limit, is not a simple task
\cite{vanNoije1,vanNoije2}. The use of the Boltzmann (or Enskog)
equation, the starting point of a systematic derivation of the
constitutive relations, is based on the \textit{Molecular Chaos}
hypothesis. This hypothesis is justified, in the dilute limit, for
an ensemble of \textit{elastic} hard spheres. Its use for
\textit{inelastic} hard spheres is, rigorously speaking, an
uncontrolled assumption. Indeed, inelasticity of the particle
collisions introduces inter-particle correlations which can
invalidate the \textit{Molecular Chaos} hypothesis. The
correlations become more pronounced as the inelasticity of the
collision increases. On the contrary, for \textit{nearly elastic}
collisions, $1-r^2 \ll 1$ (where $r$ is the coefficient of normal
restitution) the correlations are small, and the Boltzmann
equation can be used safely (again, in the dilute limit).

An additional crucial assumption, made in the process of the
derivation of the hydrodynamics from the Boltzmann equation, is
scale separation. The hydrodynamics is valid if the mean free path
of the particles is much less than any characteristic length
scale, and the mean time between two consecutive collisions is
much less than any characteristic time scale, described
hydrodynamically. This assumption (as well as the dilute gas
assumption, if used) should be verified, in every specific system,
after the hydrodynamic problem is solved and the characteristic
length and time scales determined. One more criterion for the
validity of granular hydrodynamics is the smallness of
fluctuations, driven by the noise caused by the discrete nature of
the particles. The accuracy of hydrodynamics improves with an
increase of the total number of particles in the system $N$
\cite{fluctuations}.

Once hydrodynamics is valid, it is highly rewarding using it.
Hydrodynamics has a great predictive power and helps to identify
important collective phenomena (shear flows and vortices, shocks,
different modes of clustering flows, phase separation etc.) that
are impossible to perceive in the language of individual
particles. Once identified, these collective phenomena can then be
investigated in experiment and simulations in more general (not
necessarily hydrodynamic) formulations. A recent example:
employing a variant of granular hydrodynamics, Livne \textit{et
al.} \cite{LMS1} predicted a novel phase separation instability in
a system of inelastic hard spheres, driven by a rapidly vibrating
plate at zero gravity. This system had been investigated earlier
by many workers. However, the phase separation instability, and a
plethora of interesting effects accompanying it, were overlooked.
Undoubtedly, the reason for that was that hydrodynamics had not
been fully appreciated and exploited \cite{recent}.

In this paper we employ the Navier-Stokes granular hydrodynamics
for a theoretical analysis of the \textit{process} of formation of
a density inversion in a dilute granular gas driven by a rapidly
vibrating bottom plate in a gravity field. The phenomenon of a
density inversion in the \textit{steady state} of this system is
well known in the granular community. On the contrary, the
\textit{dynamics} of the formation of the density inversion has
never been investigated previously. As we will see, this dynamics
is quite instructive. The hydrodynamic theory yields predictions
of time-dependent quantities that can be checked in molecular
dynamics simulations and experiment. Our additional motivation
here is pedagogical. While dealing with this problem, we will
present two useful hydrodynamic techniques. The first is reduction
of the order of the hydrodynamic equations, based on a hierarchy
of the length/time scales in the problem. The second technique is
the use of the Lagrangian coordinates.

Here is a layout of the remainder of this paper. In Sec. 2 we will
write down a set of hydrodynamic equations and boundary conditions
that describe the dynamics of formation of a density inversion. By
introducing scaled variables, we will delineate the two scaled
governing parameters of the problem. In Sec. 3 we briefly review
the properties of the steady state and the condition for a density
inversion. In Sec. 4 we assume a low-Mach-number flow, derive a
reduced hydrodynamic model and use it for an investigation of the
late-time dynamics of the formation of the density inversion.
Section 5 reports a numerical solution of the full hydrodynamic
equations and compares the numerical results with the predictions
of the low-Mach-number model. A discussion of the results is
presented in Sec. 6.

\section{The model problem and hydrodynamic equations}

We will adopt the simplest possible model, in two dimensions, that
exhibits a density inversion.  Let $N \gg 1$ identical nearly
elastic hard disks of diameter $d$ and mass $m=1$ move without
friction in a two-dimensional box of width $L_x$ and infinite
height. The driving base is at $y=0$, the side walls are at $x=0$
and $x=L_x$. The gravity acceleration $g$ is in the negative $y$
direction. The rapid vibrations of the base can be modeled, in a
simplified way, by prescribing a constant granular temperature
$T=T_0$ at $y=0$ (we measure the granular temperature in the units
of the velocity squared). This means that, upon collision with the
base, the particle velocity is drawn from a Maxwell's distribution
with temperature $T_0$.  The kinetic energy of the particles is
being lost by inelastic hard-core collisions parameterized by a
constant coefficient of normal restitution $r$. Particle
collisions with the side walls are assumed elastic. Alternatively,
one can specify periodic boundary conditions in the $x$-direction.

Introducing a hydrodynamic description, we deal with three
coarse-grained fields: the particle number density $n(\vec{r},t)$,
mean flow velocity $\vec{v}(\vec{r},t)$ and granular temperature
$T(\vec{r},t)$. In addition, we make two important assumptions.
First, we assume that that the collisions are nearly elastic:
$1-r^2 \ll 1$. Second, we assume that the granular gas is dilute
everywhere (including the density peak region, see below), that is
$n$ is small enough compared to the close-packing density $n_c$.
These assumptions enable us to use the classic version of the
Navier-Stokes granular hydrodynamics \cite{Jenkins}, and to limit
ourselves to the leading order terms in $n/n_c$ in the
constitutive relations. The dilute gas condition will be checked
\textit{a posteriori}.

It has been shown recently that, when this system is large enough
in the lateral direction, thermal convection may occur
\cite{Ramirez,HMD,KMcon}. In this work we will assume that $L_x$
is sufficiently small, and the convection is suppressed by the
viscosity and lateral heat conduction. Therefore, we assume flow
solely in the vertical direction. Therefore, the governing
equations are:
\begin{equation}
\frac{\partial n}{\partial t}+\frac{\partial}{\partial y}\left(n
v\right) = 0 ,\label{mass}
\end{equation}
\begin{equation}
n\left(\frac{\partial v}{\partial t}+v\frac{\partial v}{\partial
y}\right)=\frac{\partial P}{\partial y}-ng,\label{momentum}
\end{equation}
\begin{equation}
n\left(\frac{\partial T}{\partial t}+v\frac{\partial T}{\partial
y}\right) =-\frac{\partial q}{\partial y}+P\frac{\partial
v}{\partial y}-I \label{energy}.
\end{equation}
Here $P = - nT +\eta \partial v/\partial y$ is the vertical
component of the stress tensor, $q = -\kappa \partial T/\partial
y$ is the heat flux, and $I$ is the heat loss rate by the
inelastic collisions. In the nearly elastic limit, the leading
order contributions to the viscosity $\eta$ and heat conductivity
$\kappa$ are given by their classical values for for
\textit{elastic} hard disks:
$$\eta= \frac{1}{2d}\left( \frac{T}{\pi}\right)^{1/2}\,\,\,\,\mbox{and}\,\,\,\,
\kappa = \frac{2}{d}\left( \frac{T}{\pi }\right)^{1/2}\,.$$ The
heat loss rate is the following:
$$I = \pi^{1/2} (1-r^2)\,d n^2 T^{3/2}.$$
The boundary conditions for the problem include the fixed
temperature
\begin{equation}T(y=0,t) = T_0
\label{BC1}
\end{equation}
and zero velocity
\begin{equation}
v(y=0,t) = 0 \label{BC2}
\end{equation}
at the base. Now, as the total number of grains $N$ is constant:
\begin{equation}\label{conservation}
\int_0^{\infty} n(y,t) \,dy = N/L_x = const\,,
\end{equation}
the mass flux should vanish at infinity:
\begin{equation}\label{BC3}
  n v =0 \,\,\,\mbox{at}\,\,\,y = \infty.
\end{equation}
Similarly, one should assume that the momentum flux at infinity is
zero:
\begin{equation}\label{BCmomentum}
  n T + n v^2 -\eta \frac{\partial v}{\partial y} =0 \,\,\,\mbox{at}\,\,\,y =
  \infty,
\end{equation}
as well as the energy flux:
\begin{equation}
\label{BC4} nv\left(\frac{v^2}{2}+2T\right) - \eta \,v
\frac{\partial v}{\partial y} - \kappa \frac{\partial T}{\partial
y}=0 \,\,\,\mbox{at}\,\,\,y = \infty.
\end{equation}
We refer the reader to the book of Landau and Lifshitz \cite{LL}
for a detailed discussion of the divergent forms of the
hydrodynamic equations, and for the expressions for the momentum
and energy flux density used in Eqs. (\ref{BCmomentum}) and
(\ref{BC4}).

What are the characteristic length/time scales of the problem? The
fixed base temperature $T_0$ and gravity acceleration $g$ define a
characteristic macroscopic length scale $\lambda=T_{0}/g$. The
characteristic number density of the gas is therefore
$n_0=N/(\lambda L_x)$. Correspondingly, there are three
characteristic time scales in the problem. The fast time scale of
the problem, $\tau_{f} = T_0^{1/2}/g$, is independent of the
particle size.  The heat conduction time scale is $\tau =(\pi
\lambda/g)^{1/2} (Nd/L_x)$. The inelastic heat loss time scale
$\tau_l$ can be defined as $\tau_l^{-1} = (1-r^2) d \,n_0
T_0^{1/2}$.

How do we know that the time scale $\tau_f$ is indeed fast? Before
we answer this question, let us first make a preliminary
evaluation of the validity of hydrodynamics. We should demand that
the characteristic mean free path of the particles $\sim (n_0
d)^{-1}$ be much less than the macroscopic length scale $\lambda$.
This immediately follows $L_x/(Nd) \ll 1$. Notice that the
quantity $Nd/L_x$ is of the order of the number of monolayers at
rest (that is, when the system is not fluidized). Therefore, the
validity of the hydrodynamics requires that the number of
monolayers be much larger than $1$. Now we compare the
characteristic time scales $\tau_f$ and $\tau$ introduced above.
We see that the ratio $\tau_{f}/\tau$ is of the order of
$L_x/(Nd)$, which \textit{should} be much less than unity if we
want the hydrodynamics to be valid.

To make a full use of hydrodynamics, we introduce scaled
variables. Let us measure time in the units of the heat conduction
time $\tau$, the coordinate in the units of $\lambda$, the
temperature in the units of $T_0$, the density in the units of
$n_0$ and the mean flow velocity in the units of $\lambda/\tau$.
In the scaled variables, the governing equations
(\ref{mass})-(\ref{energy}) become

\begin{equation}
\frac{\partial n}{\partial t}+\frac{\partial}{\partial
y}\left(nv\right)=0 ,\label{scaledmass}
\end{equation}
\begin{equation}
\varepsilon^2 n\left(\frac{\partial v}{\partial t}+v\frac{\partial
v}{\partial y}\right)=-\frac{\partial}{\partial y}(nT)- n+
\frac{\varepsilon^2}{2}\frac{\partial}{\partial
y}\left(T^{1/2}\frac{\partial v}{\partial
y}\right),\label{scaledmomentum}
\end{equation}

\begin{equation}
 n\left(\frac{\partial T}{\partial t}+v\frac{\partial
T}{\partial y}\right)+nT\frac{\partial v}{\partial y}=
\frac{\varepsilon^2}{2} T^{1/2} \left(\frac{\partial v}{\partial
y}\right)^{2}+\frac{4}{3}\frac{\partial^{2}}{\partial
y^{2}}T^{3/2}-4\Lambda^2 n^{2}T^{3/2}, \label{scaledenergy}
\end{equation}
where
$$
\varepsilon=\frac{L_x}{\pi^{1/2} N d} \ll
1\,\,\,\,\,\text{and}\,\,\,\,\, \Lambda=\frac{(1-r^{2})^{1/2}}{2
\,\varepsilon}
$$
are the two governing scaled parameters of the problem. Parameter
$\Lambda$ is of the order of the square root of the ratio between
the heat conduction time scale and inelastic heat loss time scale.

The reduction of the four-dimensional space of governing
parameters $d,N,r$ and $L_x$ to the $(\varepsilon, \Lambda$) -
plane is an immediate benefit from using hydrodynamics. Indeed,
series of experiments or simulations with different $d,N,r$ and
$L_x$, but with the same $\varepsilon$ and $\Lambda$, should
produce the same dynamics in the scaled variables. We will see in
the following that, in the late stage of the dynamics (and in the
steady state), the only relevant parameter is $\Lambda$.

In the scaled variables, the boundary conditions (\ref{BC1}),
(\ref{BC2}), (\ref{BC3}), (\ref{BCmomentum}) and (\ref{BC4})
become $T = 1$ and $v =0$ at $y=0$; and
\begin{equation}\label{nv_scaled}
n v =0,
\end{equation}
\begin{equation}\label{BC3scaled}
n T+\varepsilon^2 n v^2 -\frac{\varepsilon^2}{2}
T^{1/2}\frac{\partial v}{\partial y}=0
\end{equation}
and
\begin{equation}\label{BC4scaled}
nv\left(\frac{\varepsilon^2}{4}
v^2+T\right)-\frac{\varepsilon^2}{4}T^{1/2}v \frac{\partial
v}{\partial y} - T^{1/2} \frac{\partial T}{\partial y}=0
\end{equation}
at $y=\infty$.

The simplest way to satisfy the boundary conditions
(\ref{BC3scaled}) and (\ref{BC4scaled}) is to assume a zero heat
flux at infinity, which implies $\partial T/\partial y = 0$ at
$y=\infty$. As the total amount of material is finite, the density
$n$ should vanish at $y=\infty$. Therefore, the $nT$ term in Eq.
(\ref{BC3scaled}) should also vanish there. Now, using what
remains of Eq. (\ref{BC3scaled}) together with Eqs.
(\ref{nv_scaled}), we see that Eq. (\ref{BC4scaled}) can be
reduced to the simple condition $\partial v/\partial y = 0$ at
$y=\infty$. Therefore, we are left with simple boundary
conditions: $T = 1$ and $v =0$ at $y=0$, and $\partial T/\partial
y=\partial v/\partial y=0$ at $y=\infty$ \cite{tempincrease}. A
full formulation of the time-dependent problem requires that we
specify the initial conditions $n(y,t=0), T(y,t=0)$ and
$v(y,t=0)$. We postpone this issue until later.

Like in many other one-dimensional gasdynamic problems \cite{ZR},
it is convenient to make a transformation of variables from the
Eulerian coordinate $y$ to the Lagrangian mass coordinate
$m(y,t)=\int_0^y n(y^{\prime},t)\,dy^{\prime}$. The physical
meaning of the Lagrangian mass coordinate is very simple: it is
the mass content of the gas on the Eulerian interval $(0, y)$ per
unit length in the $x$-direction. Notice that the infinite
interval $0\leq y<\infty$ in the Eulerian coordinates is mapped,
by this coordinate transformation, into a finite interval $0\leq
m<1$, at all times. Going over to the Lagrangian coordinate $m$,
we can rewrite Eqs. (\ref{scaledmass})-(\ref{scaledenergy}) as
follows:
\begin{equation}
\label{Lagmass} \frac{\partial}{\partial
t}\left(\frac{1}{n}\right)=\frac{\partial v}{\partial m},
\end{equation}
\begin{equation}\label{Lagmomentum}
\varepsilon^2\frac{\partial v}{\partial t}=-
\frac{\partial}{\partial m}(nT)-1
+\frac{\varepsilon^2}{2}\frac{\partial}{\partial m}\left(T^{1/2}
n\frac{\partial v}{\partial m}\right)
\end{equation}
\begin{equation}\label{Lagenergy}
\frac{\partial T}{\partial t}+nT\frac{\partial v}{\partial m} =
\frac{\varepsilon^2}{2}T^{1/2} n \left(\frac{\partial v}{\partial
m}\right)^2 + \frac{4}{3}\frac{\partial}{\partial
m}\left(n\frac{\partial}{\partial m}T^{3/2}\right)-4\Lambda^2
nT^{3/2}.
\end{equation}
The boundary conditions become $T = 1$ and $v =0$ at $m=0$, and
$n\,\partial T/\partial m=n \,\partial v/\partial m =0$ at $m=1$.

Notice that the transformation from the Eulerian to Lagrangian
coordinate involves the density $n(y,t)$ that is unknown \textit{a
priori}. We should not worry about it. After having determined the
density $n(m,t)$ in the Lagrangian coordinates, one can go back,
at any time $t$, to the Eulerian coordinate by using the relation
\begin{equation}\label{y(m)}
y(m,t) = \int_0^m \frac{dm^{\prime}}{n(m^{\prime},t)}\,.
\end{equation}

\section{Steady state profiles and density inversion}
Let us briefly review the steady state solution of the problem
\cite{HMD,KMcon,Brey2001}. To obtain the steady state profiles
$n_s(m)$ and $T_s(m)$ we simply put $\partial/\partial t = v= 0$
in Eqs. (\ref{Lagmass})-(\ref{Lagenergy}). Then Eq.
(\ref{Lagmass}) is obeyed automatically, while Eqs.
(\ref{Lagmomentum}) and (\ref{Lagenergy}) become
\begin{equation}\label{steady1}
\frac{d}{d m}\left(n_sT_s\right)+1=0
\end{equation}
and
\begin{equation}\label{steady2}
\frac{d}{d m}\left(n_s\frac{d}{d m}T_s^{3/2}\right)-3\Lambda^2
n_sT_s^{3/2}=0,
\end{equation}
respectively. As we see, the steady state is described by a single
scaled parameter $\Lambda$.  Integrating Eq. (\ref{steady1}) with
respect to $m$ and using the boundary condition at $m=1$ (that
becomes simply $nT=0$), we obtain $n_sT_s=1-m$. Now we express the
density through the temperature: $n_s(m) = (1-m)/T_s(m)$. Using
this relation in Eq. (\ref{steady2}), we arrive at the following
\textit{linear} equation \cite{HMD,KMcon,Brey2001} for
$\omega_s(m)=T_s^{1/2}(m)$:
\begin{equation}\label{steadyomega}
(1-m)\,\omega_s^{\prime\prime}-\omega_s^{\prime}-\Lambda^2(1-m)\,\omega_s=0,
\end{equation}
where primes stand for the $m$-derivatives. This is a Bessel
equation whose general solution is a linear combination of the
modified Bessel functions of the first kind
$I_{0}\left[\Lambda(1-m)\right]$ and
$K_{0}\left[\Lambda(1-m)\right]$. Demanding that the heat flux
vanish at $m=1$ and returning to the temperature variable, we
obtain:
\begin{equation}\label{steadytemp}
T_s(m)=\frac{I_{0}^2\left(\Lambda(1-m)\right)}{I_{0}^2(\Lambda)}
\end{equation}
Respectively,
\begin{equation}\label{steadydens}
n_s(m)=\frac{1-m}{T_s(m)}=\frac{(1-m) I_{0}^2(\Lambda)}
{I_{0}^2\left(\Lambda(1-m)\right)}\,.
\end{equation}

To transform back to the Eulerian coordinate $y$, we should
evaluate the integral
\begin{equation}\label{ysteady}
y(m) = \int_0^m \frac{dm^{\prime}}{n_s(m^{\prime})}
\end{equation}
numerically.  Figures \ref{BMfig1} and \ref{BMfig2} show the steady
state temperature and density profiles in the Lagrangian and
Eulerian coordinates for $\Lambda=\sqrt{10}$.

\begin{figure}
\begin{center}
\includegraphics[width=.8\textwidth]{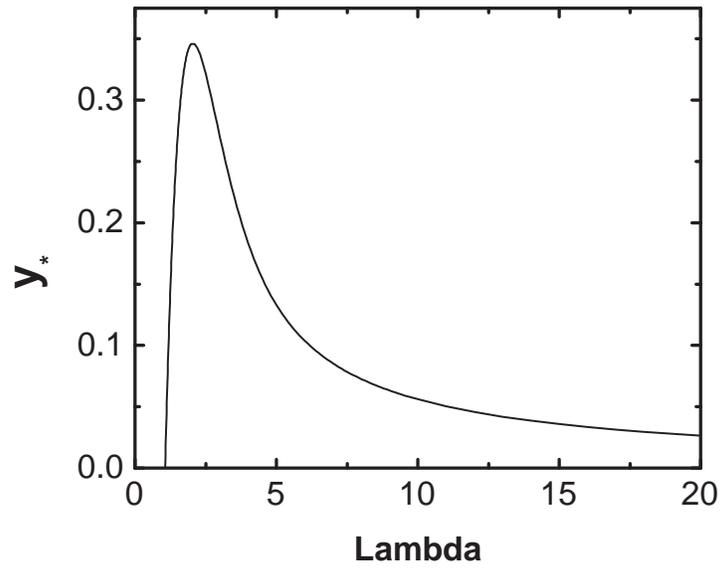}
\end{center}
\caption[]{The density peak height in the (scaled) Eulerian
coordinate versus $\Lambda$.} \label{BMmaxlocation}
\end{figure}

Equation (\ref{steadydens}) enables one to find the critical
(minimum) value of $\Lambda$ for the density inversion. At small
$\Lambda$ the density decreases monotonically with $m$ (and
therefore with $y$). The density inversion is born, at $m=0$, at
$\Lambda=\Lambda_c$. Therefore, to find $\Lambda_c$, we demand
that the derivative $n_s^{\prime}(m)$ vanish at $m=0$. Using Eq.
(\ref{steadydens}), we can reduce this condition to the equation
$I_0 (\Lambda) = 2 \Lambda\, I_1(\Lambda)$, first obtained in Ref.
\cite{Brey2001}. Its solution yields $\Lambda_c \simeq
1.06569\dots$. At $\Lambda>\Lambda_c$ the density reaches its
maximum in the Lagrangian point $m_{\ast}=1-\Lambda_c/\Lambda>0$.
Therefore, the position of the density peak in the Lagrangian
coordinate grows with $\Lambda$ monotonically. The respective
(scaled) density maximum value is
\begin{equation}\label{maximum}
 n_{max} = \frac{\Lambda_c \,I_{0}^2(\Lambda)}
{\Lambda \,I_{0}^2(\Lambda_c)}\,.
\end{equation}
Using Eq. (\ref{ysteady}), we can compute the scaled Eulerian
coordinate $y_{\ast}$ at which the maximum density maximum occurs:
\begin{equation}\label{height}
y_{\ast}=\frac{1}{I_{0}^2(\Lambda)}\int_{\Lambda_c}^{\Lambda}
\frac {I_{0}^2(\xi)} {\xi}\,d\xi\,.
\end{equation}
The dependence of $y_{\ast}$ on $\Lambda$ is non-monotonic, see
Fig. \ref{BMmaxlocation}. Very close to the density inversion
birth (at $\Lambda>\Lambda_c$) this dependence is linear:
$y_{\ast}\simeq \Lambda/\Lambda_c-1$, while at $\Lambda \gg
\Lambda_c$ one obtains $y_{\ast} \simeq (2 \Lambda)^{-1}$.
Actually, the integral in Eq. (\ref{height}) can be calculated
analytically, in terms of a hypergeometric function.

\section{Formation of a density inversion: Low-Mach-Number flow}

Now let us return to the time-dependent problem. Can we exploit
the presence of a small parameter $\varepsilon^2$ in Eqs.
(\ref{Lagmomentum}) and (\ref{Lagenergy}) to simplify the
equations? One can expect that the respective terms
\textit{cannot} be neglected during the rapid initial stage of the
dynamics. The duration of this stage is of the order of the
acoustic time or, in our scaled units, of the order of
$\varepsilon$. A comparison of different terms in the momentum
equation (\ref{Lagmomentum}) shows that the typical scaling of the
velocity during this stage is $\varepsilon^{-1} \gg 1$. In the
physical units, the flow velocity here is comparable to, or larger
than, the speed of sound. Correspondingly, at early times the
$\varepsilon^2$-terms are comparable to, or larger than, the other
terms and therefore should be kept.

The situation changes at later times, determined by the inelastic
heat losses and heat conduction. Here the characteristic Mach
number of the flow becomes small, and we can neglect, in the
leading order, the $\varepsilon^2$-terms in Eqs.
(\ref{Lagmomentum}) and (\ref{Lagenergy}). As the result, the
momentum equation (\ref{Lagmomentum}) is reduced to the
hydrostatic condition (\ref{steady1}), while in the energy
equation (\ref{Lagenergy}) one can drop the viscous heating term.
The continuity equation (\ref{Lagmass}) remains the same as
before. The boundary conditions at $m=0$ do not change, but those
at $m=1$ are reduced to $n\,\partial T/\partial m = 0$. The
boundary condition $n\,\partial v/\partial m=0$ should be dropped
altogether.

What is the physical meaning of the reduced model? A
low-Mach-number flow proceeds on a time scale that is much longer
than the time needed for establishing the (approximate)
hydrostatic equilibrium, so the flow does not violate the
hydrostatics. Similar reductions of nonlinear time-dependent
gas-dynamic problems, that employ a separation of time/length
scales, have been successfully used in the context of the dynamics
of an optically thin plasma that is cooled by radiation
\cite{Meerson}. The presence of a heat loss term in the energy
equation makes the latter problem similar to problems of granular
dynamics.

The hydrostatic condition immediately yields $p(m,t)\equiv
n(m,t)T(m,t)=1-m$, like in the steady state solution! Using it
together with the continuity equation (\ref{Lagmass}), we can
rewrite the energy equation as an evolution equation for
$\omega(m,t)=T^{1/2}(m,t)$:
\begin{equation}
\omega\frac{\partial \omega}{\partial t}=(1-m)\frac{\partial^{2}
\omega}{\partial m^{2}}- \frac{\partial \omega}{\partial
m}-\Lambda^2 (1-m)\,\omega\label{sqrtT}\,.
\end{equation}
This nonlinear parabolic equation should be solved with two
boundary conditions: $\omega(0,t)=1$ and $(1-m)\,\partial
\omega/\partial m \to 0$ as $m \to 1$. Notice that the only
relevant scaled parameter in the low-Mach-number problem is
$\Lambda$.

Once Eq. (\ref{sqrtT}) is solved, we can find $n(m,t)$ and
$v(m,t)$:
\begin{equation}\label{velocity}
v(m,t) = \frac{\partial}{\partial t} \int_0^m
\frac{dm^{\prime}}{n(m^{\prime},t)}\,.
\end{equation}
To transform back to the Eulerian coordinates we use Eq.
(\ref{y(m)}).

\begin{figure}
\begin{center}
\includegraphics[width=.7\textwidth,clip=]{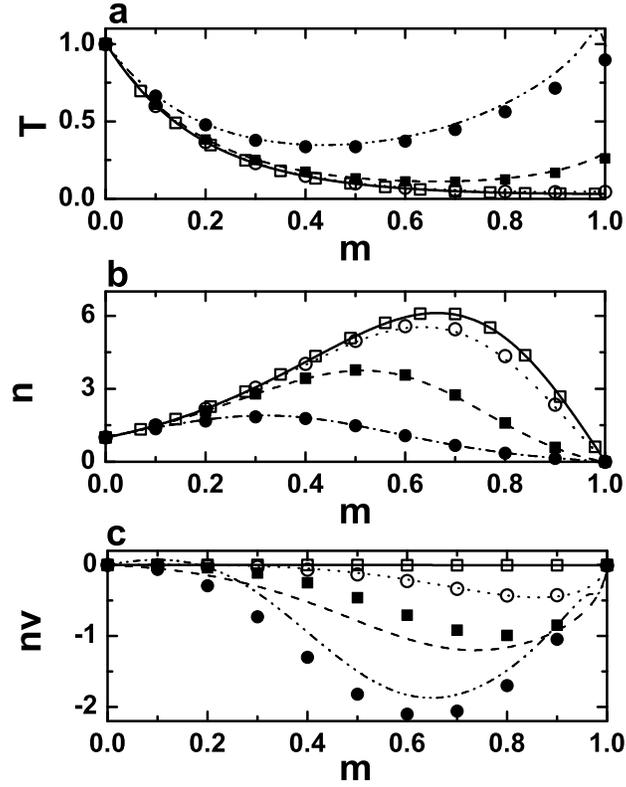}
\end{center}
\caption[]{The temperature (a), number density (b) and mass flux
$nv$ (c) versus the Lagrangian mass coordinate $m$ at late times,
as computed from the low-Mach-number model (symbols) and full
hydrodynamics (lines).  The scaled parameters are
$\Lambda=\sqrt{10}$ and $\varepsilon=\sqrt{5} \cdot 10^{-2}$. The
scaled times are $t=0.1$ (solid circle, dash-dot-dot line), $0.3$
(solid square, dashed line), $0.5$ (empty circle, dotted line) and
$0.85$ (empty square, solid line). The analytic steady state
solutions $T_s(m)$ and $n_s(m)$ are indistinguishable from the
respective solid lines.} \label{BMfig1}
\end{figure}

\begin{figure}
\begin{center}
\includegraphics[width=.8\textwidth]{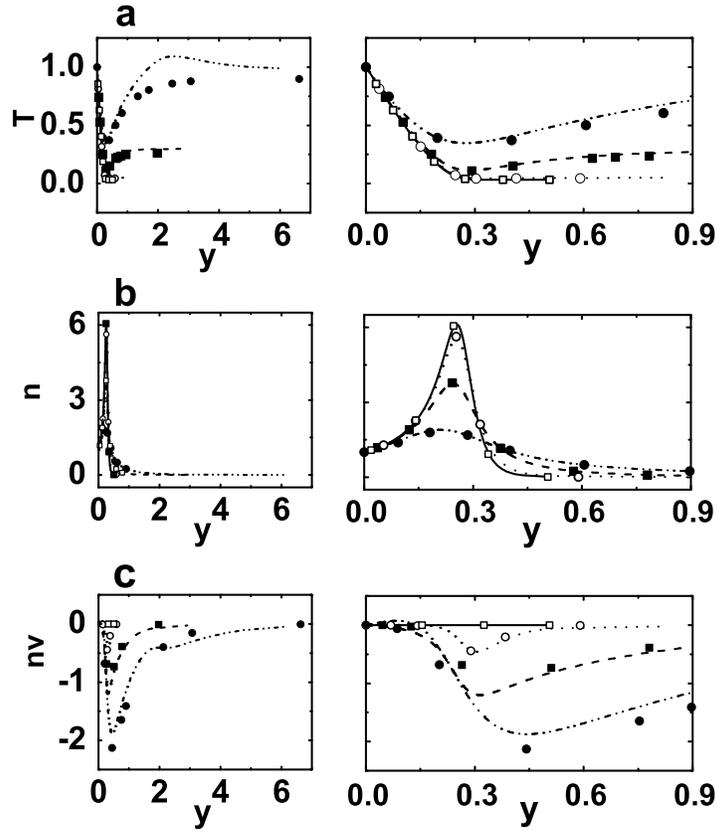}
\end{center}
\caption[]{The temperature (a), number density (b) and mass flux
$nv$ (c) versus the Eulerian coordinate $y$ at late times, as
computed from the low-Mach-number model (symbols) and full
hydrodynamics (lines).  The scaled parameters are
$\Lambda=\sqrt{10}$ and $\varepsilon=\sqrt{5} \cdot 10^{-2}$. The
figures on the left and on the right show the same profiles on
different scales of height. The scaled times are $t=0.1$ (solid
circle, dash-dot-dot line), $0.3$ (solid square, dashed line),
$0.5$ (empty circle, dotted line) and $0.85$ (empty square, solid
line). The steady state solutions $T_s(y)$ and $n_s(y)$ are
indistinguishable from the respective solid lines.} \label{BMfig2}
\end{figure}

Equation (\ref{sqrtT}) was solved numerically with a MATLAB PDE
solver. Figure~\ref{BMfig1} shows an example of numerical
solution, at different times, for $\Lambda=\sqrt{10}$. The initial
condition for the temperature, $T(m,t=0)=1$, corresponds to a
barometric density profile, which looks especially simple in the
Lagrangian coordinate: $n(m,t=0) =1-m$. The solution shows cooling
of the granular gas because of the inelastic heat losses (see Fig.
~\ref{BMfig1} a). Simultaneously, a density peak develops at an
intermediate height (Fig. ~\ref{BMfig1} b). The material flux
(Fig. ~\ref{BMfig1} c) is everywhere negative: the material slowly
falls from above until the steady-state with a zero mean velocity
is reached. An important feature of the time-dependent solution is
a transient temperature minimum that develops at intermediate
times. It is caused by the fact that the inelastic heat losses are
very small at large heights, where the gas density is small.
Therefore, the gas cools down relatively fast at intermediate
heights, which causes a temperature minimum there. Then a downward
heat flux develops that removes the heat from above, finally
creating a monotonically decreasing steady state temperature
profile. One can see from Fig. \ref{BMfig1} that the steady state
temperature and density profiles coincide with the analytic
solutions. The respective solutions in the Eulerian coordinates
are shown in Fig. \ref{BMfig2}.

\section{Formation of a density inversion: Early times}
The low-Mach-number theory is invalid at early times, when the
hydrostatics has not settled down yet, and the mean flow velocity
can reach, or even exceed, the local speed of sound. To
investigate these early times, and to check the validity of the
low-Mach-number theory at later times, we solved numerically the
full set of hydrodynamic equations
(\ref{Lagmass})-(\ref{Lagenergy}) in the Lagrangian form. The
parameter $\varepsilon$ was taken to be $\sqrt{5} \cdot 10^{-2}
\simeq 0.02236$. Equations (\ref{Lagmass}) and (\ref{Lagmomentum})
were solved with an explicit numerical scheme, based on the
standard finite-difference method described in Ref.
\cite{Richtmyer}. The viscous term in Eq. (\ref{Lagmomentum}) was
integrated explicitly, using a two-steps method, which is both
stable and accurate. The energy equation (\ref{Lagenergy}) was
treated implicitly, with a conservative scheme.

We started with simple initial conditions: the constant
temperature $T(m,t=0)=1$, the barometric density profile
$n(m,t=0)=1-m$ and a zero mean velocity. As expected, the
late-time results show good agreement with the predictions of the
low-Mach-number theory. Figure 2 shows that, at latest times, the
full-hydrodynamics solutions coincide with those obtained with the
low-Mach-number theory. At an intermediate time one can see a
local increase of the temperature at high altitudes, which is
predicted by the full hydrodynamics, but missed by the
low-Mach-number theory. The reason for this transient temperature
maximum is the relatively large viscous heating that occurs at
high altitudes because of a steep velocity gradient there.

The early-time results of the full numerical solution, transformed
back to the Eulerian coordinate, are shown in Fig. 4. One can see
that the development of the density peak, and the formation of the
transient temperature minimum, start already at early times. A
more detailed analysis does show differences between the fast and
slow theories. Indeed, a necessary condition for the validity of
the low-Mach-number theory is the hydrostatic balance condition.
This condition looks especially simple in the Lagrangian
coordinate: $p(m,t)=1-m$. That is, when the hydrostatics sets in,
the plots of the pressure versus $m$, at different times, should
collapse into a single straight line. Figure 5 shows the pressure
plots at different times. One can see that, at late times,  the
hydrostatic condition sets in as expected. At early times pressure
oscillations are observed, and hydrostatics does not hold. Figure
6 shows the pressure history at the bottom plate $m=0$, or $y=0$.
One can see that the pressure undergoes oscillations on the fast
time scale of the system $\mathcal{O} (\varepsilon)$. The
oscillations rapidly decay with time and, starting from $t\simeq
0.4$, the pressure stays almost constant which implies a
low-Mach-number flow.

\section{Discussion}
We have employed granular hydrodynamics to describe a
time-dependent granular flow that gives rise to a density
inversion in a vibrofluidized granular system. We have predicted
the formation of a transient temperature minimum at intermediate
times. The temperature minimum ultimately disappears as the system
approaches the steady state with a density inversion. We have also
predicted pressure oscillations at early times, when the
hydrostatic condition is not yet satisfied.

We hope that the predicted phenomena will be observed in molecular
dynamics (MD) simulations and in experiments with vibrofluidized
granular beds. The MD simulations should start from the barometric
density profile, corresponding to the temperature of the bottom
plate. A natural way to initiate the initial state is to first run
the simulation with \textit{elastic} particle collisions. After
the barometric profile sets in, the collision rule should be
switched into inelastic regime.

The hydrodynamic theory considered in this work is expected to be
quantitatively accurate for nearly elastic collisions, $1-r^2 \ll
1$, and in the dilute limit. The latter condition demands that the
maximum number density of the particles be less than, say, $0.2 \,
n_c$, where $n_c =2/(\sqrt{3} d^2)$ is the hexagonal close packing
density. The maximum density peak is achieved at the steady state.
Therefore, using Eq. (\ref{maximum}) and going back to the
``physical" density, we can write the dilute regime criterion as
\begin{equation}\label{dilutecritA}
  \frac{\Lambda_c}{I_{0}^2(\Lambda_c)}\,\frac{\sqrt{3} N d^2}{2 \lambda L_x}
\,\frac{I_{0}^2(\Lambda)}{\Lambda}<0.2\,.
\end{equation}
An alternative form of the same criterion is
\begin{equation}\label{dilutecritB}
  \frac{I_0^2(\Lambda)}{\varepsilon\,\Lambda}\,\frac{d}{\lambda} <
  0.65\,.
\end{equation}
We expect that the theory is still valid qualitatively when this
criterion is \textit{not} satisfied, but the maximum density is
still less than about $0.5 \,n_c$. To achieve a better accuracy
for such moderately dense flows, one can use the constitutive
relations of Jenkins and Richman \cite{Jenkins} that account for
excluded volume effects.

\subsection*{Acknowledgements} This work was supported by a grant
No. 180/02 from the Israel Science Foundation, administered by the
Israel Academy of Sciences and Humanities.

\bibliography{MEERSON}

\begin{thebibliography}{10}

\bibitem{Kadanoff}
L.~P. Kadanoff.
\newblock Built upon sand.
\newblock {\em Rev. Mod. Phys.}, 71:435, 1999.

\bibitem{clustering1}
M.~A. Hopkins and M.~Y. Louge.
\newblock Inelastic microstructure in rapid granular flows of smooth disks.
\newblock {\em Phys. Fluids A}, 3:47, 1991.

\bibitem{clustering2}
I.~Goldhirsch and G.~Zanetti.
\newblock Clustering instability in dissipative gases.
\newblock {\em Phys. Rev. Lett.}, 70:1619, 1993.

\bibitem{clustering3}
S.~McNamara and W.~R. Young.
\newblock Dynamics of a freely evolving, two-dimensional granular medium.
\newblock {\em Phys. Rev. E}, 53:5089, 1996.

\bibitem{clustering4}
J.~S. Olafsen and J.~S. Urbach.
\newblock Clustering, order, and collapse in a driven granular monolayer.
\newblock {\em Phys. Rev. Lett.}, 81:4369, 1998.

\bibitem{vanNoije1}
T.~C.~P van Noije and Ernst. M.
\newblock Kinetic theory of granular gases.
\newblock In P\"{o}schel and Luding \cite{TPSL}, page~3.

\bibitem{vanNoije2}
I.~Goldhirsch.
\newblock Granular gases: Probing the boundaries of hydrodynamics.
\newblock In P\"{o}schel and Luding \cite{TPSL}, page~79.

\bibitem{fluctuations}
It has been found recently that fluctuations are \textit{anomalously} large
  \cite{MPSS,BarratTrizac}, and their strength decreases very slowly with $N$
  \cite{MPSS}, in the parameter region corresponding to phase separation in
  driven granular gases. As the result, hydrodynamics can still be invalid in a
  system of as many as $4 \cdot 10^4$ particles \cite{MPSS}.

\bibitem{LMS1}
E.~Livne, B.~Meerson, and P.~V. Sasorov.
\newblock Symmetry-breaking instability and strongly peaked periodic clustering
  states in a driven granular gas.
\newblock {\em Phys. Rev. E}, 65:021302, 2002.

\bibitem{recent}
The phase separation instability has been recently observed in molecular
  dynamics simulations \cite{MPSS,Brey2002,Argentina}, and the subject is
  rapidly developing \cite{KM,LMS2}.

\bibitem{Jenkins}
J.~T. Jenkins and M.~W. Richman.
\newblock Kinetic theory for plane flows of a dense gas of identical, rough
  inelastic circular disks.
\newblock {\em Phys. Fluids}, 28:3485, 1985.

\bibitem{Ramirez}
R.~Ram\`{\i}rez, D.~Risso, and P.~Cordero.
\newblock Thermal convection in fluidized granular systems.
\newblock {\em Phys. Rev. Lett.}, 85:1230, 2000.

\bibitem{HMD}
X.~He, B.~Meerson, and G.~Doolen.
\newblock Hydrodynamics of thermal granular convection.
\newblock {\em Phys. Rev. E}, 65:030301(R), 2002.

\bibitem{KMcon}
E.~Khain and B.~Meerson.
\newblock Onset of thermal convection in a horizontal layer of granular gas.
\newblock {\em Phys. Rev. E}, 67:021306, 2003.

\bibitem{LL}
L.~D. Landau and E.~M. Lifshitz.
\newblock {\em Course of Theoretical Physics}, volume 6 Fluid Mechanics.
\newblock Pergamon, Oxford, 1987.

\bibitem{tempincrease}
It was observed, in some experiments \cite{Clement,Helal} and simulations
  \cite{Brey2001,Helal,RS} with vibrofluidized granular materials that, at
  large heights, the temperature starts \textit{increasing} with the height. As
  the particle density at these heights is very small, hydrodynamics might be
  inapplicable. In any case, most of the density profile is apparently
  insensitive to the presence of such a temperature inversion, if any.

\bibitem{ZR}
Y.~B. Zeldovich and Y.~P. Raizer.
\newblock {\em The Physics of Shock Waves and High Temperature Hydrodynamic
  Phenomena}.
\newblock Academic, New York, 1967.

\bibitem{Brey2001}
J.~J. Brey, M.~J. Ruiz-Montero, and F.~Moreno.
\newblock Hydrodynamics of an open vibrated granular system.
\newblock {\em Phys. Rev. E}, 63:061305, 2001.

\bibitem{Meerson}
B.~Meerson.
\newblock Nonlinear dynamics of radiative condensations in optically thin
  plasmas.
\newblock {\em Rev. Mod. Phys.}, 68:215, 1996.

\bibitem{Richtmyer}
R.~D. Richtmyer and K.~W. Morton.
\newblock {\em Difference Methods for Initial Value Problems}.
\newblock Interscience, New York, 1967.

\bibitem{TPSL}
T.~P\"{o}schel and S.~Luding, editors.
\newblock {\em Granular Gases}.
\newblock Lecture Notes in Physics. Springer, Berlin, 2001.

\bibitem{MPSS}
B.~Meerson, T.~P\"{o}schel, P.~V. Sasorov, and T.~Schwager.
\newblock Breakdown of granular hydrodynamics at a phase separation threshold.
\newblock cond-mat/020826.

\bibitem{BarratTrizac}
A.~Barrat and E.~Trizac.
\newblock A molecular dynamics ``{M}axwell demon'' experiment for granular
  mixtures.
\newblock {\em Molecular Physics (to appear)}, 2003.
\newblock cond-mat/0212054.

\bibitem{Brey2002}
J.~J. Brey, M.~J. Ruiz-Montero, F.~Moreno, and R.~Garc\'{\i}a-Rojo.
\newblock Transversal inhomogeneities in dilute vibrofluidized granular fluids.
\newblock {\em Phys. Rev. E}, 65:061302, 2002.

\bibitem{Argentina}
M.~Argentina, M.~G. Clerc, and R.~Soto.
\newblock {V}an der {W}aals-like transition in fluidized granular matter.
\newblock {\em Phys. Rev. Lett.}, 89:044301, 2002.

\bibitem{KM}
E.~Khain and B.~Meerson.
\newblock Symmetry-breaking instability in a prototypical driven granular gas.
\newblock {\em Phys. Rev. E}, 66:021306, 2002.

\bibitem{LMS2}
E.~Livne, B.~Meerson, and P.~V. Sasorov.
\newblock Symmetry breaking and coarsening of clusters in a prototypical driven
  granular gas.
\newblock {\em Phys. Rev. E}, 66:050301(R), 2002.

\bibitem{Clement}
E.~Clement and J.~Rajchenbach.
\newblock Fluidization of a bidimensional powder.
\newblock {\em Europhys. Lett.}, 16:133, 1991.

\bibitem{Helal}
K.~Helal, T.~Biben, and J.~P. Hansen.
\newblock Local fluctuations in a fluidized granular media.
\newblock {\em Physica A}, 240:361, 1997.

\bibitem{RS}
R.~Ram\`{\i}rez and R.~Soto.
\newblock Temperature inversion in granular fluids under gravity.
\newblock cond-mat/0210471.

\end{thebibliography}
\pagebreak

\pagebreak

\pagebreak

\pagebreak

\begin{figure}
\begin{center}
\includegraphics[width=.7\textwidth]{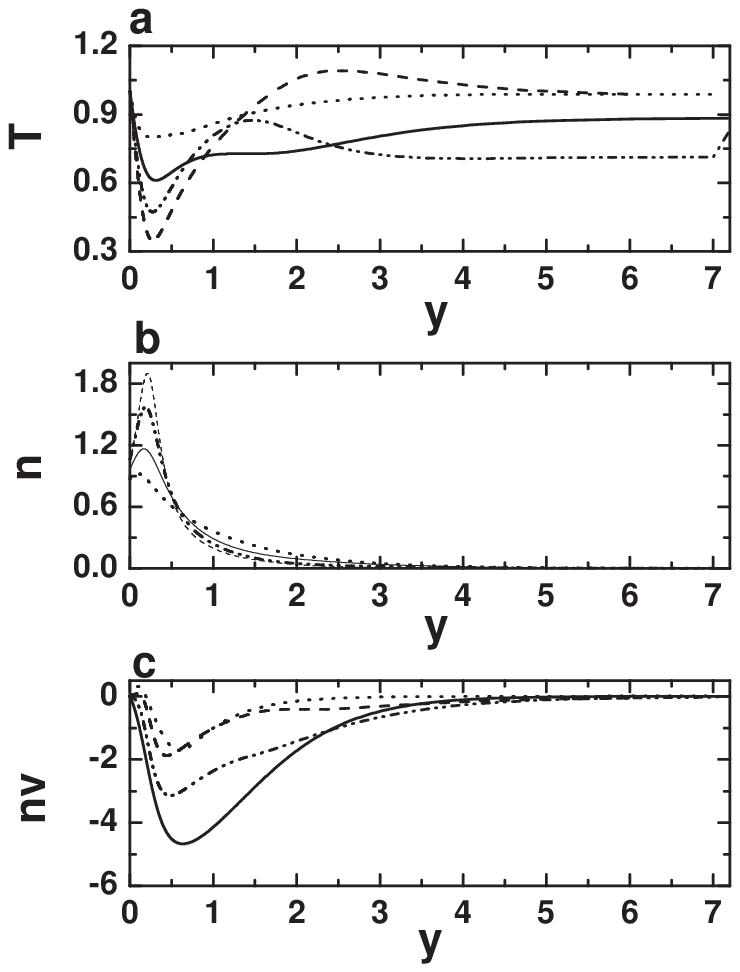}
\end{center}
\caption[]{The temperature (a), number density (b) and mass flux
$nv$ (c) versus the Eulerian coordinate $y$ at early times, as
computed from the full hydrodynamics.  The scaled parameters are
$\Lambda=\sqrt{10}$ and $\varepsilon=\sqrt{5} \cdot 10^{-2}$.  The
scaled times are $t=0.01$ (the dotted line), $0.04$ (the solid
line), $0.07$ (the dash-dot-dot line) and $0.1$ (the dashed
line).} \label{BMfig3}
\end{figure}
\pagebreak

\begin{figure}
\begin{center}
\includegraphics[width=.7\textwidth,clip=]{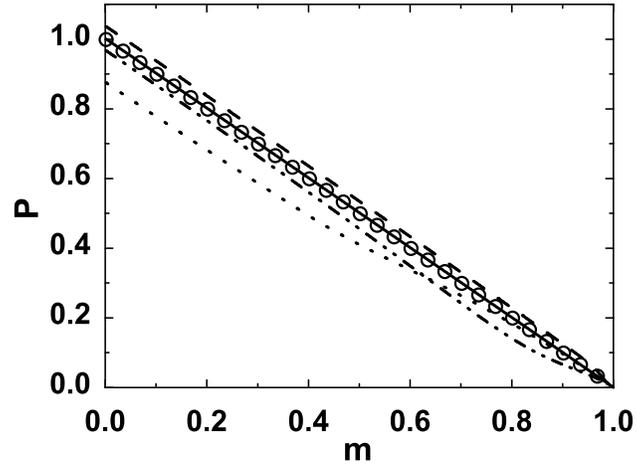}
\end{center}
\caption[]{The pressure $p=n\,T$ versus the Lagrangian mass
coordinate $m$ at different times, as computed from the full
hydrodynamics. The scaled parameters are $\Lambda=\sqrt{10}$ and
$\varepsilon=\sqrt{5} \cdot 10^{-2}$.  The scaled times are
$t=0.01$ (the dashed line), $0.04$ (the dash-dot-dot line), $0.5$
(the solid line) and $0.85$ (the empty circles).} \label{BMfig4}
\end{figure}
\pagebreak

\begin{figure}
\begin{center}
\includegraphics[width=.7\textwidth,clip=]{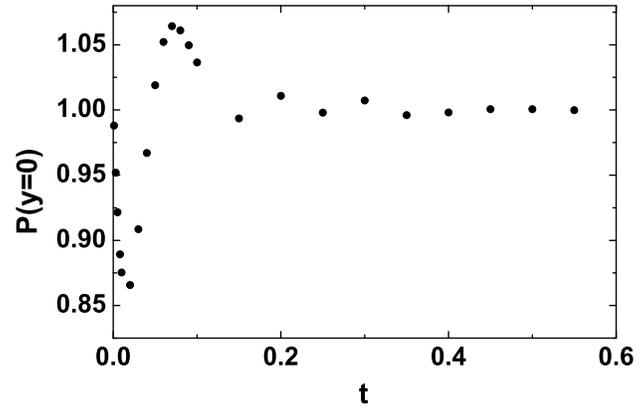}
\end{center}
\caption[]{The pressure $p=n\,T$ at the bottom plate $y=0$, versus
time, as computed from the full hydrodynamics. The scaled
parameters are $\Lambda=\sqrt{10}$ and $\varepsilon=\sqrt{5} \cdot
10^{-2}$.} \label{BMfig5}
\end{figure}
\end{document}